\documentclass[a4paper,fleqn,usenatbib]{mnras}

\usepackage{newtxtext,newtxmath}

\usepackage[T1]{fontenc}
\usepackage{ae,aecompl}

\usepackage{graphicx}	
\usepackage{amsmath}	
\usepackage{amssymb}	

\title[Low surface brightness galaxies in Perseus]{A population of faint low surface brightness galaxies in the Perseus cluster core}

\author[Wittmann et al.]{Carolin Wittmann,$^{1}$\thanks{E-mail: carolin@dwarfgalaxies.net}
Thorsten Lisker,$^{1}$
Liyualem Ambachew Tilahun,$^{1,2}$
\newauthor
Eva K. Grebel,$^{1}$
Christopher J. Conselice,$^{3}$
Samantha Penny,$^{4}$
Joachim Janz,$^{5}$
\newauthor
John S. Gallagher III,$^{6}$
Ralf Kotulla$^{6}$
and James McCormac$^{7,8}$
\\
$^{1}$Astronomisches Rechen-Institut, Zentrum f\"ur Astronomie der Universit\"at Heidelberg, M\"onchhofstra{\ss}e 12-14, 69120 Heidelberg, Germany\\
$^{2}$Department of Physics, Bahir Dar University, PO Box 79, Bahir Dar, Ethiopia\\
$^{3}$School of Physics and Astronomy, University of Nottingham, Nottingham, NG7 2RD, UK\\
$^{4}$Institute of Cosmology and Gravitation, University of Portsmouth, Dennis Sciama Building, Burnaby Road, Portsmouth PO1 3FX, UK\\
$^{5}$Centre for Astrophysics and Supercomputing, Swinburne University, Hawthorn, VIC 3122, Australia\\
$^{6}$Department of Astronomy, University of Wisconsin at Madison, 475 North Charter Street, Madison, WI 53706-1582, USA\\
$^{7}$Department of Physics, University of Warwick, Coventry CV4 7AL, UK\\
$^{8}$Isaac Newton Group of Telescopes, Apartado de correos 321 , E-38700 Santa Cruz de La Palma, Canary Islands, Spain
}

\date{Accepted 2017 May 17. Received 2017 April 20; in original form 2017 February 2}

\pubyear{2017}

\begin{document}
\label{firstpage}
\pagerange{\pageref{firstpage}--\pageref{lastpage}}
\maketitle

\begin{abstract}
We present the detection of 89 low surface brightness (LSB), and thus low stellar density galaxy candidates in the Perseus cluster core, of the kind named `ultra-diffuse galaxies', with mean effective $V$-band surface brightnesses $24.8$--$27.1$\,mag\,arcsec$^{-2}$, total $V$-band magnitudes $-11.8$ to $-15.5$\,mag, and half-light radii $0.7$--$4.1$\,kpc. The candidates have been identified in a deep mosaic covering $0.3$\,deg$^2$, based on wide-field imaging data obtained with the {\it William Herschel Telescope}. We find that the LSB galaxy population is depleted in the cluster centre and only very few LSB candidates have half-light radii larger than $3$\,kpc. This appears consistent with an estimate of their tidal radius, which does not reach beyond the stellar extent even if we assume a high dark matter content ($M/L=100$). In fact, three of our candidates seem to be associated with tidal streams, which points to their current disruption. Given that published data on faint LSB candidates in the Coma cluster -- with its comparable central density to Perseus -- show the same dearth of large objects in the core region, we conclude that these cannot survive the strong tides in the centres of massive clusters.
\end{abstract}




\section{Introduction}
\label{sec:sec1}

Galaxies of low surface brightness, once considered a rare part of the overall galaxy population \citep[e.g.,][]{vandenBergh1959}, now are recognized to exist in all galaxy mass ranges with a wide variety of properties \citep[e.g.,][]{Sprayberry1995, deBlok1996, Schombert2011, Boissier2016}. In addition, improved techniques have led to the detection of increasing numbers of low surface brightness, and thus low stellar density, galaxies \citep{Impey1996, Dalcanton1997, Kniazev2004}. These are particularly numerous among the less luminous members of galaxy clusters \citep[e.g.,][]{vanderBurg2016}.

Galaxy clusters have been and are being surveyed for increasingly faint galaxies, leading to the detection of low-mass dwarf galaxies in the surface brightness regime of Local Group dwarf spheroidals (dSphs) with mean effective surface brightnesses $\langle\mu_V\rangle_{50} > 24$\,mag~arcsec$^{-2}$, and even ultra-faint dwarfs \citep[e.g.][]{Munoz2015,Ferrarese2016}. With this increasing coverage of the parameter space of magnitude, half-light radius and surface brightness, we therefore consider it necessary to distinguish between a regular -- even though faint -- dwarf galaxy, and a low surface brightness (LSB) galaxy \emph{in the sense of having a surface brightness clearly lower than average at its luminosity}. For example, while the Virgo Cluster Catalogue of \citet{Binggeli1985} contains hundreds of newly identified dwarf galaxies, many of them being faint in magnitude and surface brightness, their catalogue also includes a handful of LSB objects that seemed to form `a new type of very large diameter (10\,000\,pc), low central surface brightness ($\geq 25\,B$\,mag\,arcsec$^{-2}$) galaxy, that comes in both early (i.e., dE) and late (i.e., Im V) types' \citep{Sandage1984}. Further Virgo cluster galaxies of dwarf stellar mass but with unusually large size and faint surface brightness were described by \citet{Impey1988}, and some similar objects were discovered in the Fornax cluster by \citet{Ferguson1988} and \citet{Bothun1991}. Three decades later, galaxies in the same general parameter range were dubbed `ultra-diffuse galaxies' by \citet{vanDokkum2015a}.

In the Coma cluster, a large number of over 700 very faint candidate member galaxies with total magnitudes $M_B > -13$\,mag, half-light radii $0.2 < r_{50} < 0.7$\,kpc and central surface brightnesses as low as $\mu_{B,0} = 27$\,mag~arcsec$^{-2}$ were identified by \citet{Adami2006}. In the brighter and overlapping magnitude range $ -11 \gtrsim M_g \gtrsim -16$\,mag \citet{vanDokkum2015a} and \citet{Koda2015} reported numerous LSB candidates with $\mu_{g,0} \geq 24$\,mag\,arcsec$^{-2}$ and half-light radii up to 5\,kpc in Coma, of which five large objects with  $r_{50} \gtrsim 3$\,kpc are spectroscopically confirmed cluster members \citep{vanDokkum2015b, Kadowaki2017}. The Virgo cluster study of \citet{Mihos2015, Mihos2017} revealed four LSB candidates with even lower central surface brightnesses of $\mu_{V,0} \sim 27$\,mag\,arcsec$^{-2}$ and half-light radii as large as 10\,kpc. In the Fornax cluster an abundant population of faint LSB galaxies with $\mu_{r,0} \geq 23$\,mag\,arcsec$^{-2}$ were catalogued by \citet{Munoz2015} and \citet{Venhola2017}, of which a few have $r_{50} > 3$\,kpc \citep{Venhola2017}. Several such objects in different environments were also reported by \citet{Dunn2010}.

Although LSB galaxies have now been detected in large numbers, their origin remains a puzzle. Especially the abundant existence of LSB galaxies of dwarf stellar mass in galaxy clusters raised the question how these low stellar density systems could survive in the tidal field of such dense environments. For example, \citet{vanDokkum2015a} did not report any signs of distortions for the faint LSB candidates identified in the Coma cluster. Other cluster LSB galaxies of dwarf luminosity harbour surprisingly large and intact globular cluster (GC) systems \citep[e.g.][]{Beasley2016b, Peng2016}. One explanation could be that these galaxies are characterized by a very high dark matter content that prevents disruption of their stellar component. A similar interpretation was given by \citet{Penny2009} for a population of remarkably round and undistorted dSphs in the Perseus cluster core. Dynamical analyses of two faint LSB galaxies in the Coma and Virgo cluster indeed revealed very high mass-to-light ratios on the order of $M/L = 50$--$100$ within one half-light radius \citep{Beasley2016a, vanDokkum2016}. Similar or even higher $M/L$ ratios are also characteristic for Local Group dSphs with $M_V > -10$\,mag or $\langle\mu_V\rangle_{50} > 25$\,mag\,arcsec$^{-2}$ \citep[cf.][]{McConnachie2012}. On the other hand, \citet{Milgrom2015} suggested that within the MOND theory high $M/L$ ratios could also be explained if the LSB galaxies would contain yet undetected cluster baryonic dark matter.

However, apparently the above does not apply to all faint cluster LSB galaxies. For example, two LSB galaxy candidates of very low stellar density in the Virgo cluster show possible signs of disruption \citep{Mihos2015, Mihos2017}. One large LSB candidate of dwarf luminosity with a very elongated shape and truncated light profile was also reported in Fornax \citep{Lisker2017}, and several further elongated large LSB candidates were described by \citet{Venhola2017}. In the Hydra I galaxy cluster, \citet{Koch2012} identified a faint LSB galaxy with S-shaped morphology, indicative of its ongoing tidal disruption. Also \citet{vanderBurg2016}, who studied populations of faint LSB candidates with $r_{50} \geq 1.5$\,kpc in eight clusters with redshifts $\mathrm{z} = 0.044$--$0.063$, reported a depletion of LSB galaxy candidates in the cluster cores, based on number counts. Similarly, the numerical simulations of \citet{Yozin2015} predict the disruption of LSB galaxies that are on orbits with very close clustercentric passages.

In this study, we aim to investigate the faint LSB galaxy population of the Perseus cluster core. Perseus is a rich galaxy cluster at a redshift of $\mathrm{z} = 0.0179$ \citep{Struble1999}. While its mass is in between the lower mass Virgo and the higher mass Coma cluster, its core reaches a density comparable to that of the Coma cluster. There are indications that Perseus is possibly more relaxed and evolved than Coma \citep[e.g.][]{Forman1982}. For example Perseus only has a single cD galaxy in its centre, while the core of Coma harbours two large galaxies. On the other hand, \citet{Andreon1994} interpreted the `non-uniform distribution of morphological types' in Perseus as an indication that this cluster is not yet virialized and instead dynamically young. This may be supported by the observation that on large scales Perseus is not a spherically symmetric cluster like Coma, but shows a projected chain of bright galaxies extending in east--west direction that is offset from the symmetric X-ray distribution.

While a significant number of regular dwarf galaxies has already been identified in a smaller field of the cluster core by \citet{Conselice2002,Conselice2003}, we focus on galaxies in the same luminosity range with $M_V > -16$\,mag (corresponding to stellar masses of $M_* \lesssim 10^8$\,M$_{\odot}$) but of fainter surface brightness and thus lower stellar density. This is made possible by our deep wide-field imaging data obtained with the 4.2\,m {\it William Herschel Telescope} ({\it WHT}) Prime Focus Imaging Platform (PFIP), reaching a $5\sigma$ $V$-band depth of about 27\,mag\,arcsec$^{-2}$. In this paper, we concentrate on LSB galaxies with $\langle\mu_V\rangle_{50} \geq 24.8$\,mag\,arcsec$^{-2}$, which corresponds to the currently often adopted surface brightness limit of $\mu_{g,0} \geq 24$\,mag\,arcsec$^{-2}$ for the so-called `ultra-diffuse galaxies'. While the definition of the latter refers to objects with $r_{50} > 1.5$\,kpc \citep[e.g.][]{vanDokkum2015a}, we will not apply any size criterion in this study and generally speak of `faint LSB galaxies', or `LSB galaxies of dwarf stellar mass'. Previous work on the low-mass galaxy population in Perseus includes also the 29 dwarf galaxies studied by \citet{Penny2009} and \citet{deRijcke2009} in {\it Hubble Space Telescope} ({\it HST}) imaging data, of which six fall within our considered surface brightness range.

This paper is organized as follows: in Section~\ref{sec:sec2}, we describe the observations, data reduction and our final mosaic. We outline the detection of the LSB sources in Section~\ref{sec:sec3}, and specify their photometry in Section~\ref{sec:sec4}. We present our results in Section~\ref{sec:sec5}, where we define our sample of LSB candidates, examine their spatial distribution in the cluster, discuss peculiar candidates and characterize their magnitude--size--surface brightness distribution in comparison to LSB candidates in the Coma cluster. We discuss our results in Section~\ref{sec:sec6}, followed by our conclusions in Section~\ref{sec:sec7}. Throughout the paper, we assume a distance of 72.3\,Mpc to the Perseus cluster with a scale of 20.32\,kpc arcmin$^{-1}$ \citep[][using the `cosmology-corrected' quantities from NED with $H_0 =  73.00$\,km\,s$^{-1}$\,Mpc$^{-1}$, $\Omega_{\mathrm{matter}} = 0.27$, $\Omega_{\mathrm{vacuum}} = 0.73$]{Struble1999}.

\section{The Data}
\label{sec:sec2}

\begin{figure*}
	\includegraphics[width=\textwidth]{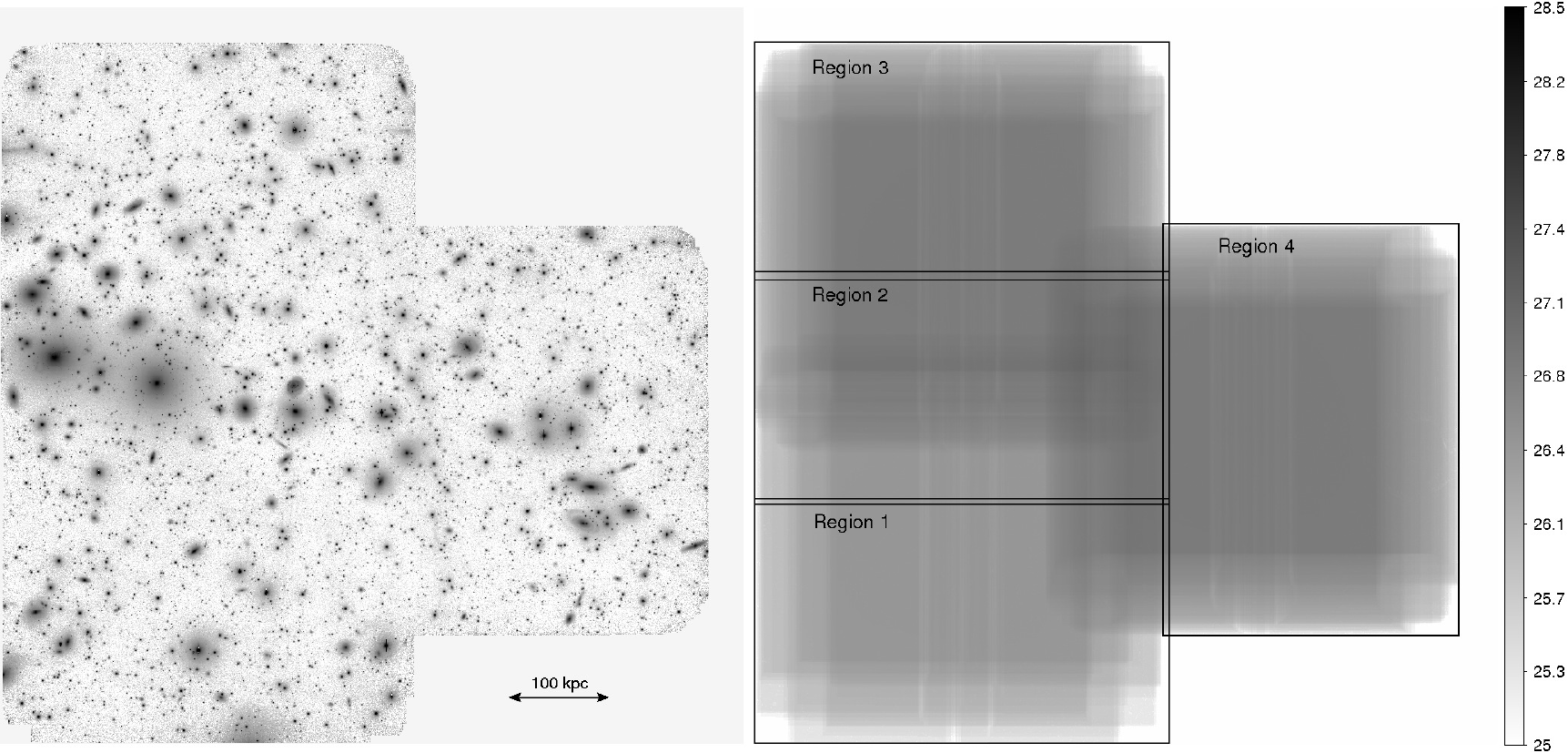}
    \caption{Deep view of the Perseus cluster core. Left: $V$-band mosaic. The image dimensions are 0.58\degr ($\hat{=}\,0.71$\,Mpc at 72.3\,Mpc) in east--west and north--south direction. North is up and east is to the left. The two bright galaxies in the east are NGC\,1275 and NGC\,1272. Right: corresponding  weight image indicating the image depth in mag\,arcsec$^{-2}$ at S/N=1 per pixel (see legend on the right-hand side). The black boxes indicate how we divided the mosaic into different regions for the detection of LSB sources (see Section~\ref{sec:sec3}).}
    \label{fig:fig1}
\end{figure*}

We acquired deep $V$-band imaging data of the Perseus cluster core with PFIP at the {\it WHT} through the Opticon programme 2012B/045 (PI T. Lisker).  The PFIP is an optical wide-field imaging camera with a field of view of $16 \times 16$\,arcmin$^2$, corresponding to $325 \times 325$\,kpc$^2$ at the distance of Perseus. The observations were carried out 2012 November 12 and 13. We performed dithered observations on three pointings across the cluster core, with individual exposure times of 120\,s. In total, 187 science exposures contribute to the final mosaic.

We reduced the data mainly with the image reduction pipeline \texttt{THELI}\footnote{\texttt{THELI GUI}, version 2.6.2} \citep{Erben2005, Schirmer2013}, which is especially designed to process wide-field imaging data. For the data reduction each exposure was spatially split into two frames, corresponding to the two detectors of the instrument. All frames were overscan- and bias-corrected, as well as flat fielded using twilight flats. To correct for remaining large-scale intensity gradients that may still be imprinted in the data after flat fielding, a master background, containing only signal from the sky, was created. For the latter the sources in all frames were masked, then the frames were normalized and stacked. Assuming the background inhomogeneities are of additive nature, the master background was subsequently subtracted from all frames. Since applying one common master background was not sufficient to remove the large-scale background variations from all frames, individual background models were created in a next step. 

The individual models are based on object-masked frames, where the masked areas were interpolated based on values from neighbouring unmasked pixels. The resulting images were convolved with a Gaussian kernel with a full width at half-maximum (FWHM) of 512 pixels. The individual background models were subtracted from each frame. We note that the applied filter kernel is large with respect to the extent of our targets, which have typical half-light radii on the order of $20$--$60$\,pixels. Then all frames were calibrated astrometrically and distortion corrected, using the Sloan Digital Sky Survey Data Release 9 (SDSS-DR9) \citep{Ahn2012} as a reference catalogue. Finally the frames were resampled and combined to a mosaic, where each frame was weighted according to the square of its inverse sky noise.

In a second iteration of the reduction we improved the individual background models of the frames that were contaminated through the extended haloes of the two brightest cluster galaxies. This optimization was done outside the \texttt{THELI} pipeline, mainly using \texttt{IRAF}.\footnote{\texttt{IRAF} is distributed by the National Optical Astronomy Observatory, which is operated by the Association of Universities for Research in Astronomy (AURA) under a cooperative agreement with the National Science Foundation.} Manually extending the masks would have resulted in a very high fraction of masked pixels on the single frames. To avoid this, we modelled the light distribution of both galaxies in the first iteration mosaic, using \texttt{IRAF} \emph{ellipse} and \emph{bmodel}. We then subtracted the galaxy models from the distortion corrected frames before generating new individual background models with \texttt{THELI}. The new background models were then subtracted from the original science frames, and combined to the second mosaic.

Lastly we corrected our mosaic for spatial zero-point variations, again outside the \texttt{THELI} pipeline. After selecting suitable stars in our mosaic using \textsc{SExtractor} \citep{Bertin1996}, we measured their magnitudes with the \texttt{IRAF} task \emph{photometry} on the individual flat fielded frames, before any background model was subtracted. We calculated the zero-point of each frame as median magnitude offset with respect to the SDSS-DR9 catalogue, using the transformation equations from \citet{Jester2005}. The zero-point variations are then given as the deviation of the magnitude offset of individual stars from the zero-point of the respective frame. We rejected stars that deviate by more than 0.2\,mag from the zero-point of the respective frame and only considered stars with small magnitude errors in both the SDSS-DR9 catalogue and the measurements with \texttt{IRAF} \emph{photometry}, requiring $\sqrt{\Delta mag^2_{phot} + \Delta mag^2_{SDSS}} < 0.05$\,mag. We then established a two-dimensional map yielding the zero-point variations across the detector by fitting a two-dimensional surface to the zero-point variations obtained for all frames. Finally, we divided each frame by this map, and repeated the above described reduction steps leading to the final mosaic. The zero-point of the final mosaic is 26 mag, with a mean variation of 0.02 mag with respect to the SDSS-DR9 catalogue.

Fig.~\ref{fig:fig1} (left-hand panel) shows our final deep mosaic of the Perseus cluster core (also Figs~\ref{fig:fig3} and~\ref{fig:fig4}). It is not centred directly on the brightest cluster galaxy NGC\,1275, but on a region including the chain of luminous galaxies that are distributed to the west of it. The mosaic covers an area of $\sim 0.27$\,deg$^2$ ($\hat{=}\,0.41$\,Mpc$^2$), and extends to a clustercentric distance of 0.57\degr ($\hat{=}\,0.70$\,Mpc$^2$) from NGC\,1275. This corresponds to 29\,per\,cent of the Perseus cluster virial radius for $R_{\mathrm{vir}} = 2.44$\,Mpc \citep{Mathews2006}, or 39\,per\,cent when adopting $R_{\mathrm{vir}} = 1.79$\,Mpc \citep{Simionescu2011}. The mosaic reaches an image depth of 27\,mag\,arcsec$^{-2}$ in the $V$-band at a signal-to-noise ratio of $\mathrm{S/N} = 1$ per pixel, with a pixel scale of 0.237 arcsec pixel$^{-1}$. The corresponding $1\sigma$ and $5\sigma$ depths are 28.6 and 26.8\,mag\,arcsec$^{-2}$, respectively. The image depth varies across the mosaic, as can be seen in the weight image (Fig.~\ref{fig:fig1}, right-hand panel). The average seeing FWHM is 0.9\,arcsec.

For the subsequent detection and photometry of low surface brightness sources we created one copy of the mosaic where we removed most of the sources with bright extended haloes, including the largest cluster galaxies and the haloes of foreground stars. We fitted the light profiles with \texttt{IRAF} \emph{ellipse}, generated models with \texttt{IRAF} \emph{bmodel} and subtracted them from the mosaic.

\section{Detection}
\label{sec:sec3}
Motivated by the detection of faint LSB galaxy candidates in the Virgo and Coma galaxy clusters by \citet{Mihos2015} and \citet{vanDokkum2015a}, we inserted LSB galaxy models in the same parameter range into our mosaic and then searched systematically for similarly looking objects in Perseus. We decided to search for LSB sources by eye, since automatic detection algorithms often fail in reliably detecting sources with very low S/N. We realized the models with a one component S\'{e}rsic profile of S\'{e}rsic index $n=0.7$--$1.2$ that were convolved with a Gaussian kernel, adopting our average seeing FWHM.

We generated a first set of 27 models in the parameter range $24.6 \leq \langle\mu_V\rangle_{50} \leq 27.8$\,mag\,arcsec$^{-2}$, $-14 \geq M_V \geq -16.6$\,mag, and $2.1 \leq r_{50} \leq 9.7$\,kpc, assuming an average foreground extinction of $A_V = 0.5$\,mag at the location of Perseus. Among them are nine model types with different magnitudes and half-light radii. For each model type we generated two additional variants with altered position angle and ellipticity, which results in slightly different surface brightnesses. We created a second set of seven nearly round (ellipticity = 0.1) models with $\langle\mu_V\rangle_{50} \leq 26.0$\,mag\,arcsec$^{-2}$ that extend the parameter range to smaller half-light radii of 1.5\,kpc and fainter magnitudes of $-13.5$\,mag.

From the first model set, we always inserted $30$--$40$ models of one type, i.e. with the same magnitude and half-light radius but varying ellipticity, into one copy of the mosaic. We generated two additional mosaic copies where we inserted the models from the second model set. We used these copies only at a later stage to focus the detection especially on smaller and fainter LSB sources that turned out to be quite numerous based on the search using the first model set. In total we inserted 305 models from the first model set into nine different mosaic copies, and 56 models from the second set into two further copies.

To facilitate the visual detection of LSB sources, we used the mosaic variant where we previously fitted and subtracted the light distribution of most of the extended sources (see Section~\ref{sec:sec2}). To remove the remaining bright sources on each copy of the mosaic, we ran \textsc{SExtractor} to detect all sources with more than 10 pixels above a detection threshold of 1.5\,$\sigma$, and replaced the pixels above this threshold with zero values, corresponding to the background level of our mosaic. We then convolved the data with a circular Gaussian kernel with $\sigma=1$\,pixel, and demagnified each copy by a factor of 1.5. We further divided each mosaic copy into four smaller regions of different image depth according to the weight image (see Fig.\ref{fig:fig1}, right-hand panel). Finally two of us independently searched visually for diffuse sources in each copy, thereby detecting simultaneously the inserted models and real LSB candidates, without knowing where the former had been inserted. After removing sources that we identified more than once in different copies of the same region, this resulted in a preliminary sample of 214 LSB sources that were identified by at least one of us, and for which we carried out photometry (see Section~\ref{sec:sec4}).

\begin{figure}
	\includegraphics[width=\columnwidth]{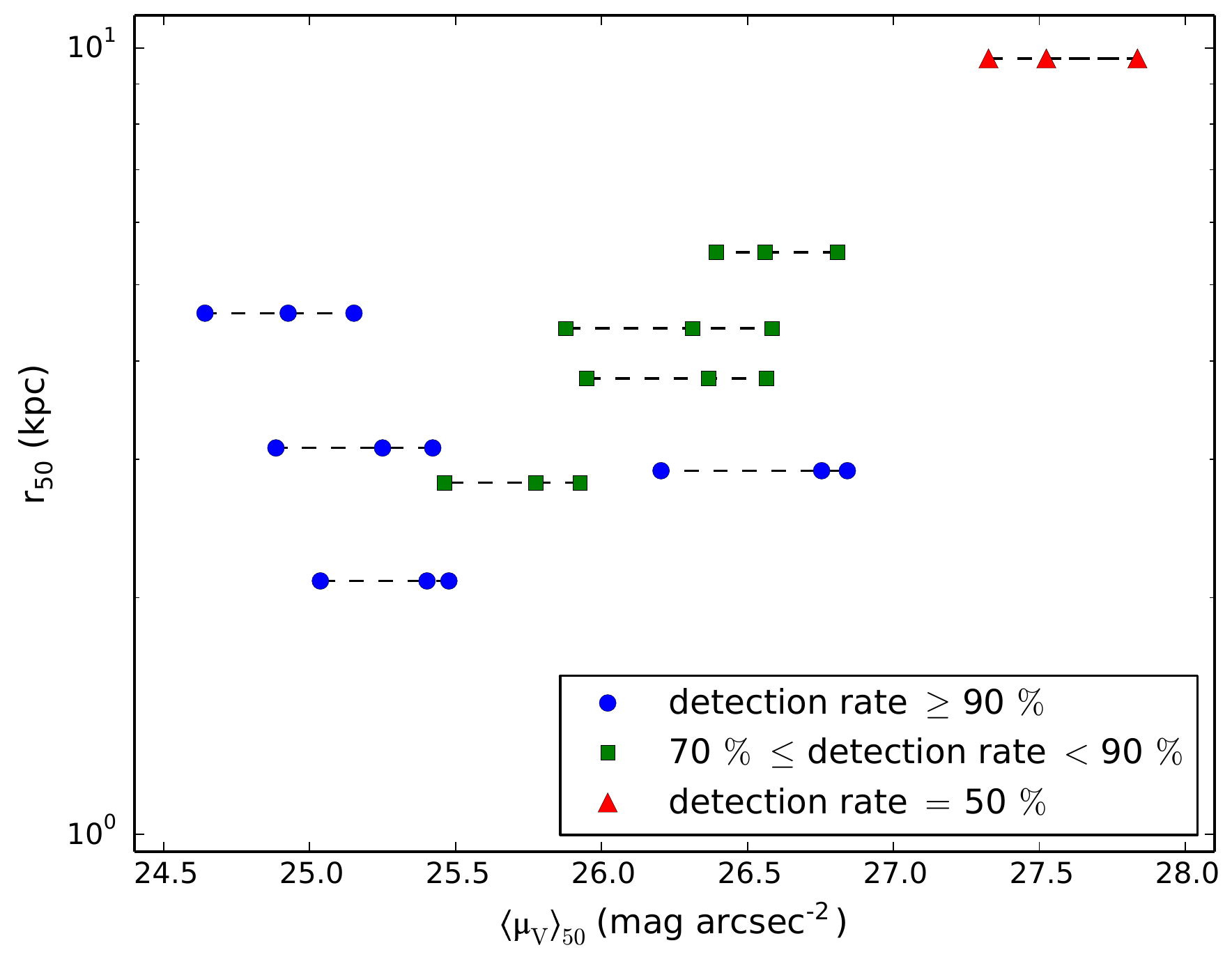}
    \caption{Detection rates of visually identified model galaxies as a function of half-light radius and surface brightness. The detection rates are based on $30$--$40$ models of one type, with the same half-light radius and magnitude, but varying ellipticity and thus surface brightness, that were inserted into one copy of the mosaic, respectively. Models of the same type are connected through dashed lines in the plot. The total number of inserted models is 305.}
    \label{fig:fig2}
\end{figure}

We used the visually identified models from the first model set to get a rough estimate on our detection rate (see Fig.~\ref{fig:fig2}). We estimated the detection rate for each model type as fraction of the total number of inserted models that were visually identified. We find that the detection rate generally drops with surface brightness. We detected more than 90\,per\,cent of all models with $\langle\mu_V\rangle_{50} < 25.5$\,mag\,arcsec$^{-2}$, between 70 and 90\,per\,cent of all models with $25.5 \leq \langle\mu_V\rangle_{50} < 27.0$\,mag\,arcsec$^{-2}$, and about $50$\,per\,cent of all models with $\langle\mu_V\rangle_{50} > 27.0$\,mag\,arcsec$^{-2}$.\footnote{The given surface brightnesses refer to the average surface brightness of the three model variants with different ellipticity, and thus surface brightness, that exist per model type.}

The models with $\langle\mu_V\rangle_{50} < 27.0$\,mag\,arcsec$^{-2}$ are in general clearly visible in our data and the main reason for missing some of them seems to be related to overlap with brighter sources. We estimated the area occupied by remaining bright extended sources in our object-subtracted mosaic to be 12\,per\,cent\footnote{This accounts for all sources that were detected with \textsc{SExtractor} with more than 1000 connected pixels above a detection threshold of $1.5\,\sigma$.}, which compares to an average detection rate of 90\,per\,cent of all models with $\langle\mu_V\rangle_{50} < 27.0$\,mag\,arcsec$^{-2}$. Scatter in the trend of decreasing detection fraction with surface brightness can both be caused by our approach of visual source detection, as well as by the different overlap fractions of the inserted models with brighter sources.\footnote{We note that the fraction of models whose centre overlaps with one of the \textsc{SExtractor}-detected sources above $1.5\,\sigma$ does not exceed 12\,per\,cent per model type.} The detection rate of models with $\langle\mu_V\rangle_{50} < 27.0$\,mag\,arcsec$^{-2}$ is similar in all regions of our mosaic, even in the shallowest region ({\it Region 1}; see Fig.\ref{fig:fig1}, right-hand panel). For models with $\langle\mu_V\rangle_{50} > 27.0$\,mag\,arcsec$^{-2}$ we find, however, a lower detection rate in {\it Region 1} and {\it Region 2}, compared to the other two regions. While {\it Region 1} is the shallowest region, the lower detection rate in {\it Region 2} might be related to the higher galaxy density compared to the other regions.

\section{Photometry}
\label{sec:sec4}

Photometry of LSB sources is challenging and the measurements suffer in general from higher uncertainties compared to sources of brighter surface brightness. One reason for this is that the radial flux profile of the former is characterized by a larger fraction of flux at large radii, where the S/N is typically very low. This also implies that contamination from close neighbour sources and the presence of background gradients is more severe for these objects. We quantify the arising uncertainties in our data using inserted LSB galaxy models (see Section~\ref{sec:sec5.3}).

We derived magnitudes and sizes from growth curves through iterative ellipse fitting with \texttt{IRAF} \emph{ellipse}, rather than from fits to analytical models. The first step was to obtain a first guess of the centre, ellipticity and position angle of all sources. We used \textsc{SExtractor} to measure the parameters of 131 objects that were detected with a detection threshold of $1\,\sigma$ (128 objects) or $0.8\,\sigma$ (3 objects). For 83 objects that were not detected with \textsc{SExtractor} or that had obviously wrong parameters we estimated their centre and shape visually based on the Gaussian smoothed and demagnified mosaic. Then we ran \emph{ellipse} with fixed parameters, adopting the previously measured or estimated centres, ellipticities and position angles. We chose a linear step-size of 5 pixels for consecutive isophotes. We used the first ellipse fit results to generate two-dimensional brightness models with \texttt{IRAF} \emph{bmodel} that we subtracted from the fitted source.

The residual images served as a basis to create masks of neighbouring sources from \textsc{SExtractor} segmentation images. We ran \textsc{SExtractor} in two passes, one with a minimum number of 28 connected pixels above a detection threshold of $1\,\sigma$, the other with a lower detection threshold of $0.6\,\sigma$ and requiring a minimum number of 1000 connected pixels. In both passes, we used \textsc{SExtractor} with the built-in filtering prior to detection. We combined both segmentation images and extended the masked areas by smoothing with a Gaussian kernel. We ran \emph{ellipse} in a second pass with the masks to exclude that flux from neighbouring sources contributes to the ellipse fits. From the second iteration residual images we created improved masks where the masked regions are somewhat larger. We unmasked the centre of nucleated candidates and ellipse fit residuals when necessary.

The next step was to determine the background level from the third pass ellipse fit results using the improved masks. Getting the background level right is a very subtle task and the major source of the uncertainties in the magnitude and size measurements. Therefore, we determined the background level for each of our detected LSB objects individually. We first measured the radial flux profiles out to large radii (350\,pixels) for each object. We then manually adjusted the radius and width of the background annulus, whose median flux we adopted as the background level. The inner radius of the background annulus was set at the first break in the flux profile where the intensity gradient significantly changes and the flux profile levels out. We set the width of the annulus to 50 pixels. Its shape follows the ellipticity and position angle of the measured object.

Although all neighbour sources were carefully masked, still some flux profiles show signs of contamination. Especially at larger radii where faint flux levels are reached, the flux of the LSB source can be comparable to the flux of a neighbour source that still extends beyond the masked area (e.g. some very extended haloes of foreground stars or bright cluster galaxies). Also background inhomogeneities remaining in the data after the reduction can contaminate the flux profiles. Possible contamination can become apparent in a flux profile when, for example, the profile continues to decline after the first break instead of levelling out to zero. In this case we nevertheless set the inner radius of the background annulus to the first break in the profile, and eventually decrease its width to make sure that the flux profile is flat in this region.

Even though we might truncate a galaxy at too high intensity, resulting in a systematically fainter magnitude and a smaller half-light radius, restricting the analysis to the uncontaminated inner profile helps to preserve the true surface brightnesses (see the right-hand panels in Fig.~\ref{fig:fig6} and Section~\ref{sec:sec5.3}). After subtracting the background offset, we then obtained a first estimate of the magnitudes and sizes by running \emph{ellipse} in a fourth pass on the background corrected images and taking into account the masked sources. We determined the total flux from the cumulative flux profile\footnote{We adopted the median of the cumulative fluxes \texttt{TFLUX\_E} from the ellipse fit tables, namely of the five isophotes between the inner radius of the background annulus and 20\,pixels further, as an estimate of the total flux. Since \emph{ellipse} does not account for masked regions when calculating the total flux within an isophote, we replaced the masked regions with values from the 2-D model created with \texttt{IRAF} \emph{bmodel} from the radial flux profile.} and derived the half-light radius along the semimajor axis, as well as the mean effective surface brightness within one half-light radius.

In the final iteration we measured the centre, ellipticity and position angle of our LSB sources more accurately, using our first guess parameters as input values. We used \texttt{IRAF} \emph{imcentroid} to derive the centre, and calculated the ellipticity and position angle from the image moments within a circular area defined by our first-guess half-light radius. We also further improved the masks by manually enlarging the masks of extended neighbour sources with faint haloes.\footnote{Using {\texttt{SAOImages DS9}} \citep{Joye2003} \emph{regions} and \texttt{IRAF} \emph{mskregions}.} After that we ran \emph{ellipse} in a fifth pass with the new parameters and masks to adjust the inner radius of the background annulus. We adopted the new background level and derived the final magnitudes, half-light radii and mean effective surface brightnesses in a last pass of ellipse fitting. We corrected the derived magnitudes for extinction, using the IRSA Galactic Reddening and Extinction Calculator, with reddening maps from \citet{Schlafly2011}. The average foreground extinction of our measured sources is $A_V = 0.5$\,mag.

\section{Faint LSB galaxies in the Perseus cluster core}
\label{sec:sec5}
\vspace{0.01cm}
\subsection{Sample}
\label{sec:sec5.1}

\begin{figure*}
    \includegraphics[width=\textwidth]{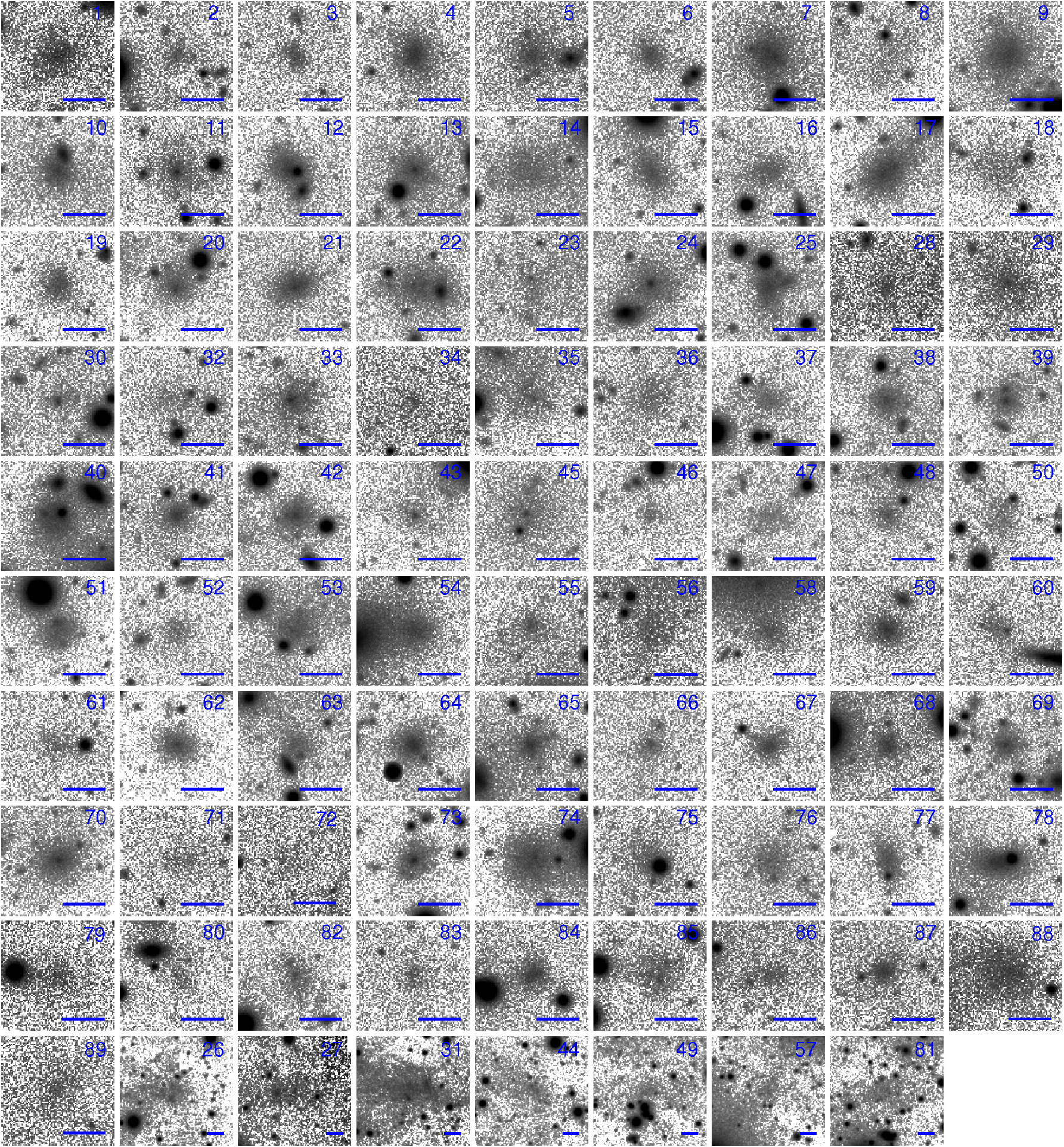}
    \caption{Sample of faint LSB galaxy candidates in the Perseus cluster core. The first 82 panels show the LSB candidates in cutout regions of our original data with a size of $21 \times 21$\,arcsec$^2$, respectively. The seven last panels in the bottom row show the LSB candidates classified as possible galaxies (see Section~\ref{sec:sec5.1}). They are displayed in our smoothed and demagnified data in cutout regions with a size of $53 \times 53$\,arcsec$^2$, respectively. The blue bar in each panel of the figure denotes a length of 3\,kpc. The number in each panel corresponds to the ID of the shown object given in Table~\ref{tab:table1}. North is up and east is to the left.}
    \label{fig:fig3}
\end{figure*}

We define our sample of LSB galaxy candidates to include all objects with $\langle\mu_V\rangle_{50} \geq 24.8$\,mag\,arcsec$^{-2}$. This corresponds to the currently often adopted surface brightness limit of $\mu_{g,0} \geq 24.0$\,mag\,arcsec$^{-2}$ for `ultra-diffuse galaxies' \citep[e.g.][]{vanDokkum2015a}, when assuming an exponential profile with S\'{e}rsic $n=1$ \citep[cf.][]{Graham2005}, $g-r = 0.6$ and using the transformation equations from \citet{Jester2005}. Of our preliminary sample, 133 objects fall into this parameter range. We carefully examined all of them, both on the original as well as on the smoothed and demagnified mosaic. We also compared them to an independent data set of the Perseus cluster, obtained with WIYN/ODI in the $g$,$r$ and $i$ filters (programme 15B-0808/5, PI: J. S. Gallagher). Since the single-band images are shallower than our data, we used the stacked $g$,$r$,$i$ images for the comparison.

Based on a more detailed visual examination of their morphology, we classified 82 of our candidates as likely galaxies. They are characterized by a smooth morphology and are confirmed in the independent data set. We classified seven further candidates as possible galaxies (all of them are shown in Fig.~\ref{fig:fig3} in the bottom row). Three of them (candidates 26, 31 and 44) are clearly visible in our data, but their morphology does not appear very regular. Since these objects are also visible in the WIYN/ODI data, we rule out that they are image artefacts. However a confusion with cirrus cannot be excluded (see Section~\ref{sec:sec5.3}). The four other candidates (candidates 27, 49, 57 and 81) are classified as possible galaxies since they are only barely visible in our data, due to their low surface brightness or low S/N, and are not confirmed in the shallower independent data set. We rejected 44 LSB sources from our sample, since we cannot exclude that these are remaining background inhomogeneities from the reduction, or residuals from ellipse fitting of the brighter galaxies. Most of them are of very diffuse nature (80\,per~cent have $\langle\mu_V\rangle_{50} \geq 26.5$\,mag\,arcsec$^{-2}$) and often do not have a smooth morphology.

Our final sample includes 89 LSB galaxy candidates in the Perseus cluster core. We show our sample in Fig.~\ref{fig:fig3} and provide the photometric parameters in Table~\ref{tab:table1}. We also compare our sample to overlapping {\it HST}/ACS images, in order to investigate whether some of our objects would classify as background sources, based on possible substructure in the form of, e.g., spiral arms. Seven of our LSB candidates fall on {\it HST}/ACS pointings, and none of them shows signs of substructure. We therefore expect that the overall contamination through background galaxies is low in our sample, based on the morphological appearance in the {\it HST} as well as in the {\it WHT} images and due to the location of our sample in the core region rather than in the cluster outskirts. Certain cluster membership can, however, only be established through measurements of radial velocities. The six brightest candidates in the {\it HST}/ACS images with $24.8 \leq \langle\mu_V\rangle_{50} \leq 25.4$\,mag\,arcsec$^{-2}$, as measured in our data,  were previously identified in \citet[][]{Penny2009} (candidates 62, 64, 69, 70, 73 and 87). One of them (candidate 62) was first catalogued by \citet{Conselice2002,Conselice2003}. The faintest candidate, with $\langle\mu_V\rangle_{50} = 26.5$\,mag\,arcsec$^{-2}$ (candidate 82), is only barely visible in the {\it HST}/ACS images and was not published previously.

\begin{table*}
	\centering
	\caption{Coordinates and structure parameters of faint LSB galaxy candidates in the Perseus cluster core. $M_V$ and $\langle\mu_V\rangle_{50}$ are corrected for Galactic foreground extinction. $A_V$ is derived from the reddening maps of \citet{Schlafly2011}. No reliable individual errors can be provided, but the right-hand panels in Fig.~\ref{fig:fig6} illustrate the statistical and systematic uncertainties for the LSB galaxy models in the parameter range of our sample; details are provided in Section~\ref{sec:sec5.3}. The table is sorted by increasing right ascension.}
	\label{tab:table1}
	\begin{tabular}{cccccccc}
		\hline
		ID & R.A. & Dec. & $\langle\mu_V\rangle_{50}$ & $M_V$ & $A_V$ & $r_{50}$ & Ellipticity\\
	       & (J2000) & (J2000) & (mag\,arcsec$^{-2}$) & (mag) & (mag) & (kpc) & \\
		\hline
\phantom{0}1 & 03 17 00.37 & +41 19 20.6 & 24.9 & -15.0 & 0.4 & 1.9 & 0.08\\ 
\phantom{0}2 & 03 17 03.26 & +41 20 29.1 & 25.9 & -12.9 & 0.4 & 1.2 & 0.20\\ 
\phantom{0}3 & 03 17 04.42 & +41 30 39.2 & 25.2 & -12.7 & 0.4 & 0.8 & 0.17\\ 
\phantom{0}4 & 03 17 07.13 & +41 22 52.5 & 25.2 & -14.5 & 0.4 & 1.7 & 0.08\\ 
\phantom{0}5 & 03 17 11.02 & +41 34 03.3 & 25.3 & -14.3 & 0.4 & 1.7 & 0.13\\ 
\phantom{0}6 & 03 17 13.29 & +41 22 07.6 & 25.3 & -12.9 & 0.4 & 0.9 & 0.10\\ 
\phantom{0}7 & 03 17 15.97 & +41 20 11.7 & 25.1 & -15.1 & 0.4 & 2.1 & 0.05\\ 
\phantom{0}8 & 03 17 19.71 & +41 34 32.5 & 26.3 & -13.7 & 0.4 & 2.1 & 0.21\\ 
\phantom{0}9 & 03 17 23.50 & +41 31 40.1 & 25.1 & -14.2 & 0.4 & 1.4 & 0.01\\ 
10 & 03 17 24.94 & +41 26 09.7 & 25.1 & -13.6 & 0.4 & 1.1 & 0.17\\ 
11 & 03 17 35.49 & +41 18 12.7 & 25.2 & -13.6 & 0.4 & 1.1 & 0.05\\ 
12 & 03 17 36.78 & +41 23 01.6 & 25.2 & -14.0 & 0.4 & 1.4 & 0.09\\ 
13 & 03 17 38.21 & +41 31 56.9 & 25.1 & -13.6 & 0.4 & 1.1 & 0.13\\ 
14 & 03 17 39.22 & +41 31 03.5 & 25.9 & -13.9 & 0.4 & 1.7 & 0.09\\ 
15 & 03 17 39.42 & +41 24 45.0 & 25.5 & -13.7 & 0.4 & 1.3 & 0.13\\ 
16 & 03 17 41.79 & +41 24 01.9 & 25.8 & -13.2 & 0.4 & 1.2 & 0.12\\ 
17 & 03 17 44.16 & +41 21 18.4 & 25.0 & -14.4 & 0.4 & 1.5 & 0.15\\ 
18 & 03 17 48.34 & +41 18 38.9 & 25.9 & -14.1 & 0.4 & 2.0 & 0.13\\ 
19 & 03 17 53.17 & +41 19 31.9 & 25.5 & -13.9 & 0.4 & 1.4 & 0.03\\ 
20 & 03 17 54.66 & +41 24 58.8 & 25.2 & -13.3 & 0.4 & 1.0 & 0.07\\ 
21 & 03 18 00.81 & +41 22 23.0 & 24.9 & -13.6 & 0.4 & 1.0 & 0.11\\ 
22 & 03 18 05.55 & +41 27 42.4 & 25.8 & -14.2 & 0.5 & 2.1 & 0.25\\ 
23 & 03 18 09.55 & +41 20 33.5 & 26.4 & -12.2 & 0.5 & 1.0 & 0.12\\ 
24 & 03 18 13.08 & +41 32 08.3 & 25.3 & -13.8 & 0.5 & 1.3 & 0.11\\ 
25 & 03 18 15.44 & +41 28 35.2 & 24.9 & -13.4 & 0.5 & 0.9 & 0.17\\ 
26 & 03 18 19.50 & +41 19 24.8 & 26.5 & -13.8 & 0.5 & 2.3 & 0.15\\ 
27 & 03 18 20.79 & +41 45 29.3 & 26.3 & -14.0 & 0.4 & 2.3 & 0.14\\ 
28 & 03 18 21.66 & +41 45 27.6 & 25.9 & -13.9 & 0.4 & 1.8 & 0.13\\ 
29 & 03 18 23.33 & +41 45 00.6 & 25.6 & -14.7 & 0.4 & 2.2 & 0.04\\ 
30 & 03 18 23.40 & +41 36 07.7 & 25.6 & -12.3 & 0.5 & 0.7 & 0.08\\ 
31 & 03 18 24.32 & +41 17 30.7 & 26.0 & -15.5 & 0.5 & 4.1 & 0.17\\ 
32 & 03 18 24.46 & +41 18 28.4 & 26.5 & -13.0 & 0.5 & 1.5 & 0.09\\ 
33 & 03 18 25.86 & +41 41 06.9 & 25.5 & -14.0 & 0.5 & 1.5 & 0.06\\ 
34 & 03 18 26.92 & +41 14 09.5 & 25.7 & -12.4 & 0.5 & 0.8 & 0.03\\ 
35 & 03 18 28.18 & +41 39 48.5 & 25.8 & -13.9 & 0.5 & 1.9 & 0.21\\ 
36 & 03 18 29.19 & +41 41 38.9 & 26.2 & -13.1 & 0.5 & 1.4 & 0.04\\ 
37 & 03 18 30.36 & +41 22 29.8 & 25.9 & -12.1 & 0.5 & 0.8 & 0.13\\ 
38 & 03 18 32.11 & +41 27 51.5 & 25.4 & -13.1 & 0.5 & 0.9 & 0.05\\ 
39 & 03 18 32.13 & +41 32 12.3 & 25.2 & -12.8 & 0.5 & 0.8 & 0.19\\ 
40 & 03 18 33.25 & +41 40 56.1 & 25.2 & -13.9 & 0.5 & 1.3 & 0.12\\ 
41 & 03 18 33.57 & +41 41 58.3 & 25.2 & -13.4 & 0.5 & 1.0 & 0.06\\ 
42 & 03 18 33.60 & +41 27 45.5 & 25.1 & -13.5 & 0.5 & 1.0 & 0.04\\ 
43 & 03 18 34.57 & +41 24 18.6 & 26.1 & -12.9 & 0.5 & 1.3 & 0.19\\ 
44 & 03 18 34.73 & +41 22 40.5 & 27.1 & -13.6 & 0.5 & 2.6 & 0.09\\ 
45 & 03 18 36.14 & +41 21 59.4 & 26.2 & -13.9 & 0.5 & 2.2 & 0.22\\ 
46 & 03 18 37.51 & +41 24 16.0 & 26.3 & -11.8 & 0.5 & 0.8 & 0.03\\ 
47 & 03 18 38.96 & +41 30 06.8 & 26.6 & -12.8 & 0.5 & 1.5 & 0.13\\ 
48 & 03 18 39.53 & +41 39 30.4 & 25.8 & -12.6 & 0.5 & 1.0 & 0.20\\ 
49 & 03 18 39.84 & +41 38 58.4 & 27.1 & -12.7 & 0.5 & 1.9 & 0.26\\ 
50 & 03 18 39.92 & +41 20 09.0 & 26.3 & -13.2 & 0.5 & 1.5 & 0.11\\ 
51 & 03 18 41.38 & +41 34 01.3 & 25.5 & -13.7 & 0.5 & 1.5 & 0.27\\ 
52 & 03 18 42.60 & +41 38 33.0 & 26.1 & -12.3 & 0.5 & 0.9 & 0.04\\ 
53 & 03 18 44.65 & +41 34 07.7 & 25.4 & -13.5 & 0.5 & 1.2 & 0.09\\ 
54 & 03 18 44.95 & +41 24 20.4 & 24.9 & -13.9 & 0.5 & 1.1 & 0.11\\ 
55 & 03 18 46.16 & +41 24 37.1 & 26.2 & -14.3 & 0.5 & 2.4 & 0.09\\ 
56 & 03 18 48.02 & +41 14 02.4 & 25.9 & -14.3 & 0.5 & 2.3 & 0.23\\ 
57 & 03 18 48.43 & +41 40 35.1 & 27.1 & -13.3 & 0.5 & 2.4 & 0.11\\ 
58 & 03 18 50.74 & +41 23 09.1 & 25.4 & -13.0 & 0.4 & 1.0 & 0.17\\ 
59 & 03 18 54.32 & +41 15 29.2 & 24.9 & -14.0 & 0.5 & 1.1 & 0.02\\ 
		\hline
	\end{tabular}
\end{table*}

\begin{table*}
	\centering
	\contcaption{}
	\label{tab:continued}
	\begin{tabular}{cccccccc}
		\hline
		ID & R.A. & Dec. & $\langle\mu_V\rangle_{50}$ & $M_V$ & $A_V$ & $r_{50}$ & Ellipticity\\
	       & (J2000) & (J2000) & (mag\,arcsec$^{-2}$) & (mag) & (mag) & (kpc) & \\
		\hline
60 & 03 18 55.38 & +41 17 50.0 & 25.8 & -12.5 & 0.5 & 1.0 & 0.18\\ 
61 & 03 18 59.40 & +41 25 15.4 & 26.0 & -12.5 & 0.4 & 1.0 & 0.07\\ 
62 & 03 18 59.42 & +41 31 18.7 & 25.4 & -13.9 & 0.4 & 1.4 & 0.07\\ 
63 & 03 19 01.50 & +41 38 59.0 & 25.8 & -12.9 & 0.5 & 1.1 & 0.17\\ 
64 & 03 19 05.83 & +41 32 34.4 & 24.8 & -13.8 & 0.4 & 1.1 & 0.09\\ 
65 & 03 19 07.77 & +41 27 12.1 & 24.8 & -12.9 & 0.4 & 0.7 & 0.06\\ 
66 & 03 19 09.32 & +41 41 51.7 & 25.9 & -12.5 & 0.5 & 0.9 & 0.06\\ 
67 & 03 19 12.76 & +41 43 30.0 & 25.2 & -13.5 & 0.5 & 1.1 & 0.08\\ 
68 & 03 19 15.01 & +41 22 31.7 & 25.1 & -13.3 & 0.4 & 0.9 & 0.06\\ 
69 & 03 19 15.70 & +41 30 34.6 & 25.1 & -12.9 & 0.4 & 0.8 & 0.05\\ 
70 & 03 19 15.86 & +41 31 05.8 & 25.2 & -14.2 & 0.4 & 1.4 & 0.03\\ 
71 & 03 19 16.02 & +41 45 45.9 & 26.1 & -13.3 & 0.5 & 1.4 & 0.05\\ 
72 & 03 19 17.53 & +41 12 41.3 & 26.7 & -12.8 & 0.4 & 1.5 & 0.02\\ 
73 & 03 19 17.83 & +41 33 48.4 & 24.9 & -13.7 & 0.4 & 1.0 & 0.07\\ 
74 & 03 19 21.94 & +41 27 22.5 & 24.9 & -14.7 & 0.4 & 1.7 & 0.15\\ 
75 & 03 19 23.06 & +41 23 16.8 & 26.3 & -13.7 & 0.4 & 2.1 & 0.20\\ 
76 & 03 19 23.12 & +41 38 58.7 & 26.0 & -13.4 & 0.5 & 1.5 & 0.11\\ 
77 & 03 19 32.76 & +41 36 12.8 & 25.7 & -13.6 & 0.4 & 1.4 & 0.09\\ 
78 & 03 19 33.80 & +41 36 32.5 & 24.8 & -13.6 & 0.5 & 1.1 & 0.34\\ 
79 & 03 19 39.19 & +41 12 05.6 & 25.4 & -14.4 & 0.4 & 1.8 & 0.06\\ 
80 & 03 19 39.22 & +41 13 43.5 & 26.3 & -12.8 & 0.4 & 1.3 & 0.07\\ 
81 & 03 19 44.03 & +41 39 18.4 & 26.9 & -13.8 & 0.4 & 2.7 & 0.14\\ 
82 & 03 19 45.66 & +41 28 07.3 & 26.1 & -13.9 & 0.4 & 2.0 & 0.13\\ 
83 & 03 19 47.45 & +41 44 09.3 & 26.0 & -12.9 & 0.4 & 1.2 & 0.07\\ 
84 & 03 19 49.70 & +41 43 42.6 & 24.8 & -13.5 & 0.4 & 0.9 & 0.05\\ 
85 & 03 19 50.13 & +41 24 56.3 & 25.5 & -13.7 & 0.4 & 1.3 & 0.05\\ 
86 & 03 19 50.56 & +41 15 33.4 & 25.6 & -12.1 & 0.4 & 0.7 & 0.17\\ 
87 & 03 19 57.41 & +41 29 31.2 & 25.0 & -13.3 & 0.4 & 0.9 & 0.05\\ 
88 & 03 19 59.10 & +41 18 33.1 & 24.8 & -15.5 & 0.4 & 2.2 & 0.02\\ 
89 & 03 20 00.20 & +41 17 05.1 & 25.7 & -13.5 & 0.4 & 1.4 & 0.10\\ 
		\hline
	\end{tabular}
\end{table*}

\subsection{Properties}
\label{sec:sec5.2}

\begin{figure*}
	\includegraphics[width=\textwidth]{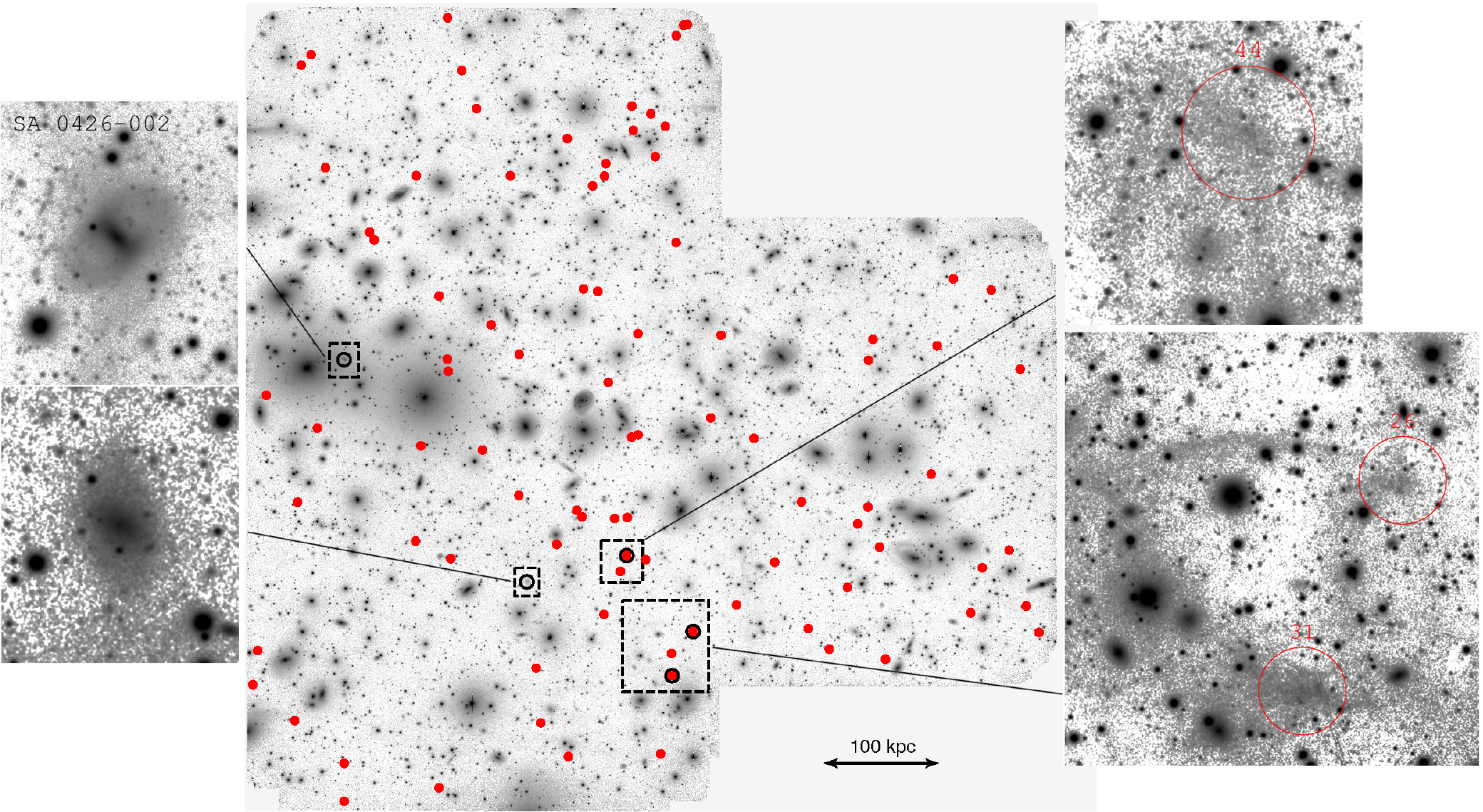}
    \caption{Spatial distribution of LSB galaxy candidates in the Perseus cluster core (central panel) and candidates with signs of possible tidal disruption (side panels). Red dots indicate our sample of LSB candidates. The dashed squares on the mosaic indicate the size of the cutout regions shown as side panels. These images were smoothed except the image in the top left side panel, which  shows a cutout from the original data. The red dots with black circles mark the positions of candidates 26, 31 and 44 shown in the side panels on the right-hand side. The two galaxies with the tidal structures in the left side panels are not part of our LSB galaxy sample. The image height and width of the mosaic is 0.58\,deg ($\hat{=}\,0.71$\,Mpc). North is up and east is to the left. }
    \label{fig:fig4}
\end{figure*}

\begin{figure}
	\includegraphics[width=\columnwidth]{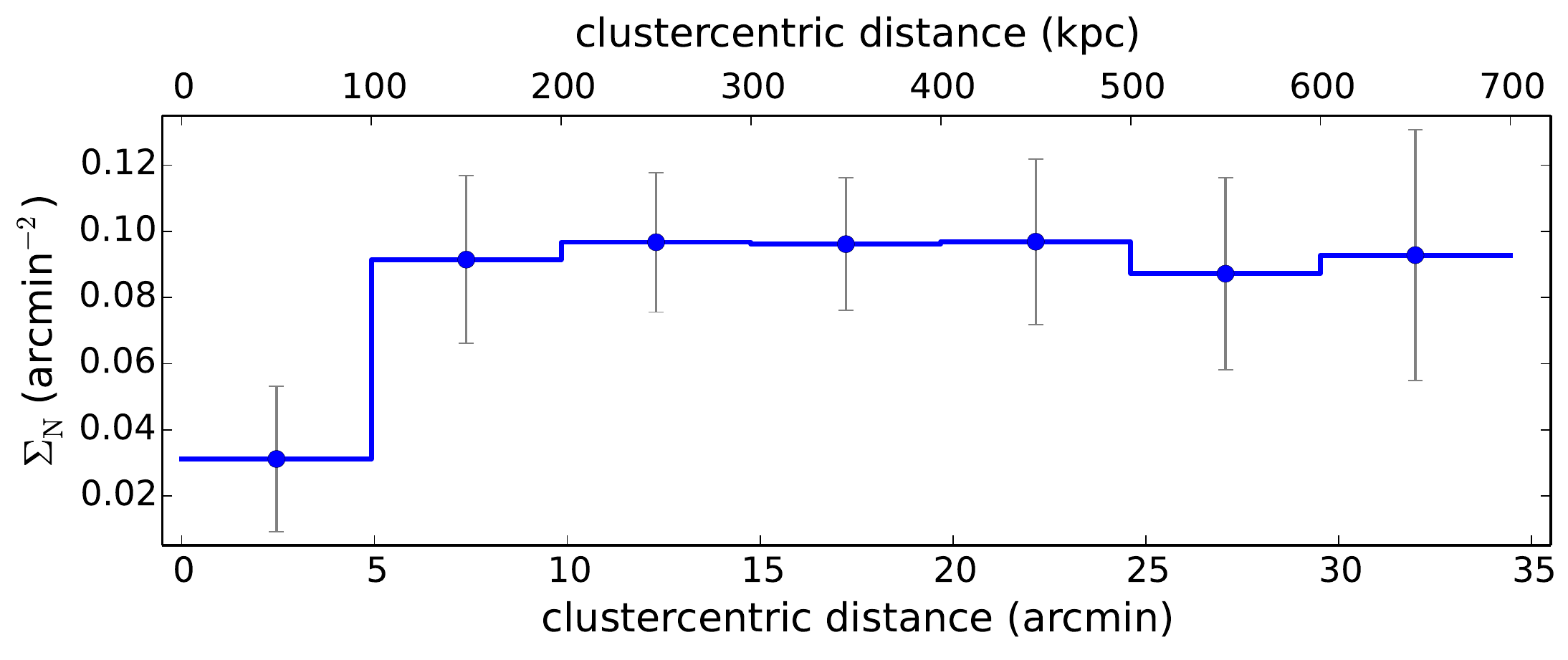}
    \caption{Radial projected number density distribution as a function of clustercentric distance of our sample of LSB galaxy candidates in the Perseus cluster core. The radial bins have a width of 100\,kpc. Shown are the statistical error bars.}
    \label{fig:fig5}
\end{figure}

\begin{figure*}
	\includegraphics[width=\textwidth]{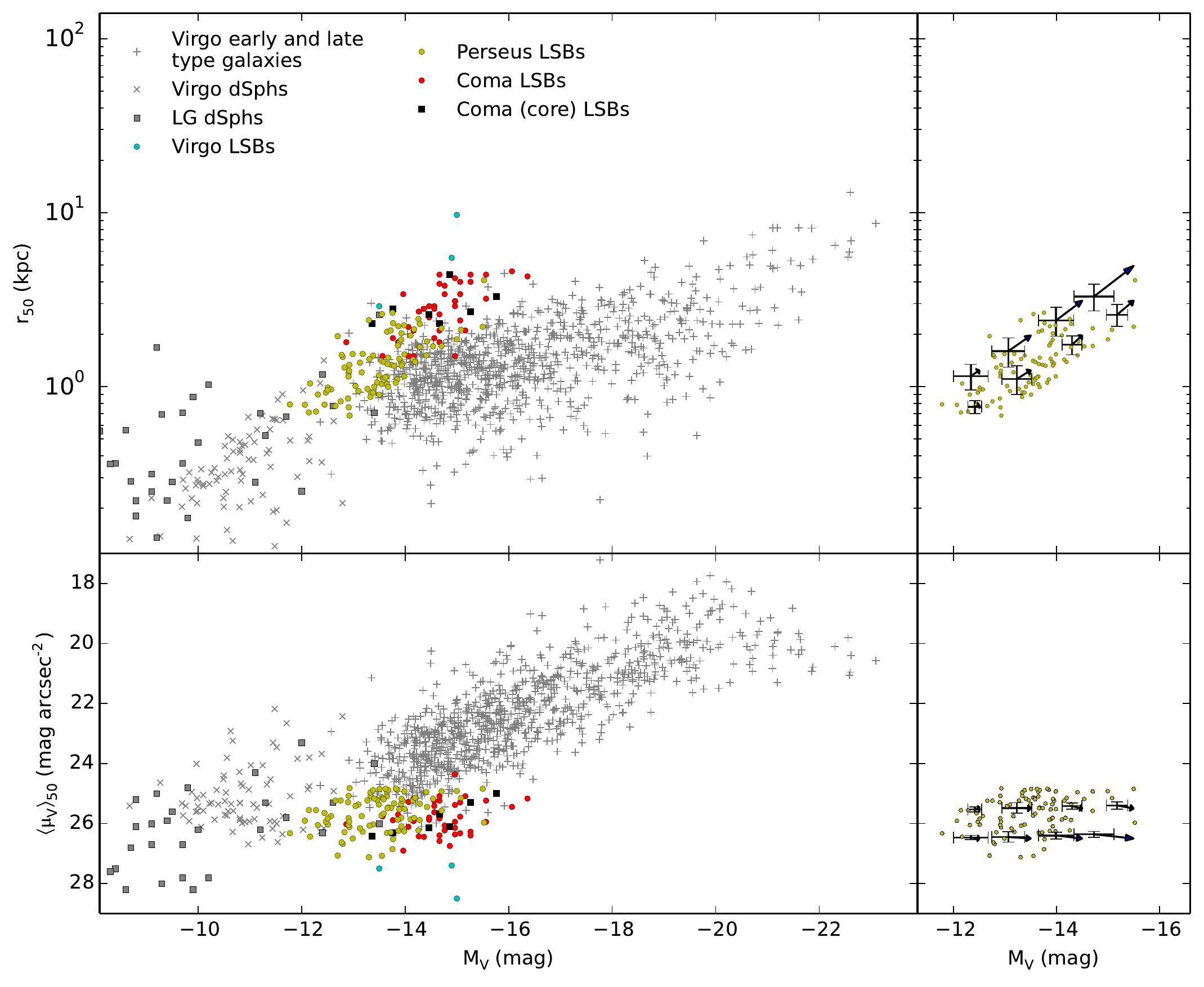}
    \caption{Structural parameters of faint LSB galaxy candidates in the Perseus cluster core (left-hand panels). We compare our sample to LSB galaxy candidates in the Virgo cluster \citep{Mihos2015}, and to LSB galaxies and candidates in the Coma cluster \citep{vanDokkum2015a}. We mark those LSB objects in Coma that are located in the cluster core within a circular area with a radius of $R = 0.15~R_{vir,Coma}$. This corresponds to an area of similar extent as our observed area of the Perseus cluster core (see Section~\ref{sec:sec5.2}). For comparison, we also show early- and late-type galaxies from the Virgo cluster (compilation of \citealt{Lisker2013}; based on the VCC), Virgo dSphs \citep{Lieder2012} and dSphs from the Local Group \citep{McConnachie2012}. We transformed the magnitudes of the LSB candidates from \citet{vanDokkum2015a} and the galaxies from \citet{Lisker2013} with the transformation equations from \citet{Jester2005}. For the former, we assumed $g-r = 0.6$, for the latter we used the measured $g-r$ colours. The two panels on the right-hand side show our typical uncertainties that occur for LSB galaxy models ($n=1$, ellipticity = 0.1) in the parameter range of our sample (see Section~\ref{sec:sec5.3}). We created eight model types with different parameters. Each model type was inserted 10 times at different positions into one copy of our mosaic. The black arrows indicate our systematic parameter uncertainties. The arrow tips point to the true parameters of the models, the endpoints represent the average measured parameter values of the 10 inserted models of each type. On average the measured $M_V$ values are by 0.4\,mag too faint, the measured $r_{50}$ values are underestimated by 0.5\,kpc and the measured $\langle\mu_V\rangle_{50}$ values are by 0.1\,mag\,arcsec$^{-2}$ too bright. The error bars represent our statistical uncertainties, and were calculated as standard deviation of the measured values of each model type.}
    \label{fig:fig6}
\end{figure*}

Fig.~\ref{fig:fig4} shows the spatial distribution of our sample of 89 faint LSB galaxy candidates in the Perseus cluster core. The sample spans a range of $47 \leq d \leq 678$\,kpc in projected clustercentric distance, with respect to the cluster's X-ray centre\footnote{The X-ray centroid almost coincides with the optical location of NGC\,1275.} \citep{Piffaretti2011}. This corresponds to $0.02-0.28$\,$R_{\mathrm{vir}}$ when assuming a virial radius of $R_{\mathrm{vir}}=2.44$\,Mpc \citep{Mathews2006}. About half of our sample is located closer than 330\,kpc to the cluster centre.

We find three LSB candidates that appear to be associated with structures resembling tidal streams (see Fig.~\ref{fig:fig4}, right-hand panels). Candidate 44 seems to be embedded in diffuse filaments, candidates 26 and 31 appear connected via an arc-shaped stream. We find one further galaxy with tidal tails (see Fig.~\ref{fig:fig4}, bottom left panel), which has a slightly brighter surface brightness of $\langle\mu_V\rangle_{50} = 24.4$\,mag\,arcsec$^{-2}$ and therefore was not included in our sample. We will analyse faint cluster galaxies with brighter surface brightnesses in a future paper. It is noticeable that all four objects are confined within one region to the south--west of the cluster centre, within a clustercentric distance range of about $300 - 400$\,kpc. Also the peculiar more luminous galaxy SA~0426-002 \citep[cf.][]{Conselice2002,Penny2014} falls on our mosaic, which shows a disturbed morphology with extended low surface brightness lobes (see Fig.~\ref{fig:fig4}, top left panel).

We show the radial projected number density distribution of our sample in Fig.~\ref{fig:fig5}. It was derived by dividing the number of galaxies in radial bins of a width of 100\,kpc by the area of the respective bin that falls on our mosaic. The bins are centred on the Perseus X-ray centre. We find that the number density is nearly constant for clustercentric distances $r \geq 100$\,kpc, but drops in the very centre at $r < 100$\,kpc,\footnote{Only two galaxies are contained in the central bin with $r < 100$\,kpc.} with a statistical significance of $2.8\,\sigma$ with respect to the average number density at larger radii. For comparison, a preliminary analysis showed that the distribution of bright cluster members is consistent with the expectation of being much more centrally concentrated.

Fig.~\ref{fig:fig6} shows the magnitude--size and magnitude--surface brightness distribution of our Perseus cluster LSB galaxy sample. We include the Coma cluster LSB galaxies and candidates from \citet{vanDokkum2015a} and the three very low surface brightness galaxy candidates in Virgo from \citet{Mihos2015}. For comparison, we also show Virgo cluster early- and late-type galaxies (compilation of \citealt{Lisker2013}; based on the Virgo Cluster Catalogue (VCC), \citealt{Binggeli1985}), Virgo cluster dSphs \citep{Lieder2012}, as well as dSphs from the Local Group \citep{McConnachie2012}.

Our sample spans a parameter range of $24.8 \leq \langle\mu_V\rangle_{50} \leq 27.1$\,mag\,arcsec$^{-2}$, $-11.8 \geq M_V \geq -15.5$\,mag and $0.7 \leq r_{50} \leq 4.1$\,kpc. The surface brightness range of our sample is comparable to the LSB galaxy sample from \citet{vanDokkum2015a} and approaches the surface brightness of the two brighter Virgo LSB candidates from \citet{Mihos2015}. With regard to magnitudes and sizes our sample includes smaller and fainter LSB candidates than the sample from \citet{vanDokkum2015a}, which is likely due to their resolution limit. At faint magnitudes, our samples overlaps with the parameter range of cluster and Local Group dSphs. We note that the apparent relation between magnitude and size of our sample is created artificially. The bright surface brightness limit arises due to our definition of including only sources fainter than $\langle\mu_V\rangle_{50} = 24.8$\,mag\,arcsec$^{-2}$ in our sample. The faint limit is due to our detection limit. 

At brighter magnitudes $M_V \leq -14$\,mag, the LSB candidates of our sample are systematically smaller at a given magnitude than the LSB candidates identified in the Coma cluster, with all but one LSB candidate having $r_{50} < 3$\,kpc. However, \citet{vanDokkum2015a} cover a much larger area of the Coma cluster, while we only surveyed the core region of Perseus.\footnote{According to tests with the inserted model galaxies (see Section~\ref{sec:sec3}) sources in the surface brightness range of the LSB galaxy sample from \citet{vanDokkum2015a} can easily be detected in our data.} Our total observed area corresponds to $0.41$\,Mpc$^2$. This translates to a circular equivalent area with a radius of $R = 0.15$\,$R_{\mathrm{vir,Perseus}}$, when assuming a virial radius for Perseus of $R_{\mathrm{vir,Perseus}} = 2.44$\,Mpc \citep{Mathews2006}.\footnote{We note that our field is not centred directly on the cluster centre, but extends to the west of it.} 

When selecting all LSB candidates from the \citet{vanDokkum2015a} sample that are located in the core of Coma, within a circular area with clustercentric distances smaller than $R = 0.15$\,$R_{\mathrm{vir,Coma}}$, where $R_{\mathrm{vir,Coma}} = 2.8$\,Mpc \citep{Lokas2003}, seven LSB candidates remain. These are marked with black squares in Fig.~\ref{fig:fig6}. One can see that also only two of them reach sizes of $ r_{50} > 3$\,kpc. Since the sample of \citet{vanDokkum2015a} has a brighter magnitude and larger size limit than our study, we restrict the comparison to objects with $M_V \leq -14$\,mag and $r_{50} \geq 2$\,kpc, which should well have been detected by \citet{vanDokkum2015a}. Five LSB candidates in the Coma cluster core are in this parameter range, whereas in Perseus we find seven. A similar result is obtained when comparing to the independent sample of Coma cluster LSB galaxy candidates from \citet{Yagi2016}. When selecting LSB candidates of the Coma core region in the same surface brightness range as our sample and with $M_V \leq -14$\,mag and $r_{50} \geq 2$\,kpc, we find 10 LSB candidates in this parameter range, where three LSB candidates have $r_{50} \geq 3$\,kpc. While it seems that the Virgo cluster galaxies shown in Fig.~\ref{fig:fig6} are also rare in this parameter range, we note that the catalogue we used is not complete at magnitudes fainter than $M_r = -15.2$\,mag.

Thus, in summary, we find that first, the core regions of the Perseus and the Coma cluster harbour a similar number of faint LSB galaxy candidates in the same parameter range of $M_V \leq -14$\,mag and $r_{50} \geq 2$\,kpc, and secondly, that large LSB candidates with $r_{50} \geq 3$\,kpc seem to be very rare in both cluster cores.

\subsection{Uncertainties}
\label{sec:sec5.3}

\begin{figure}
\centering
	\includegraphics[width=\columnwidth]{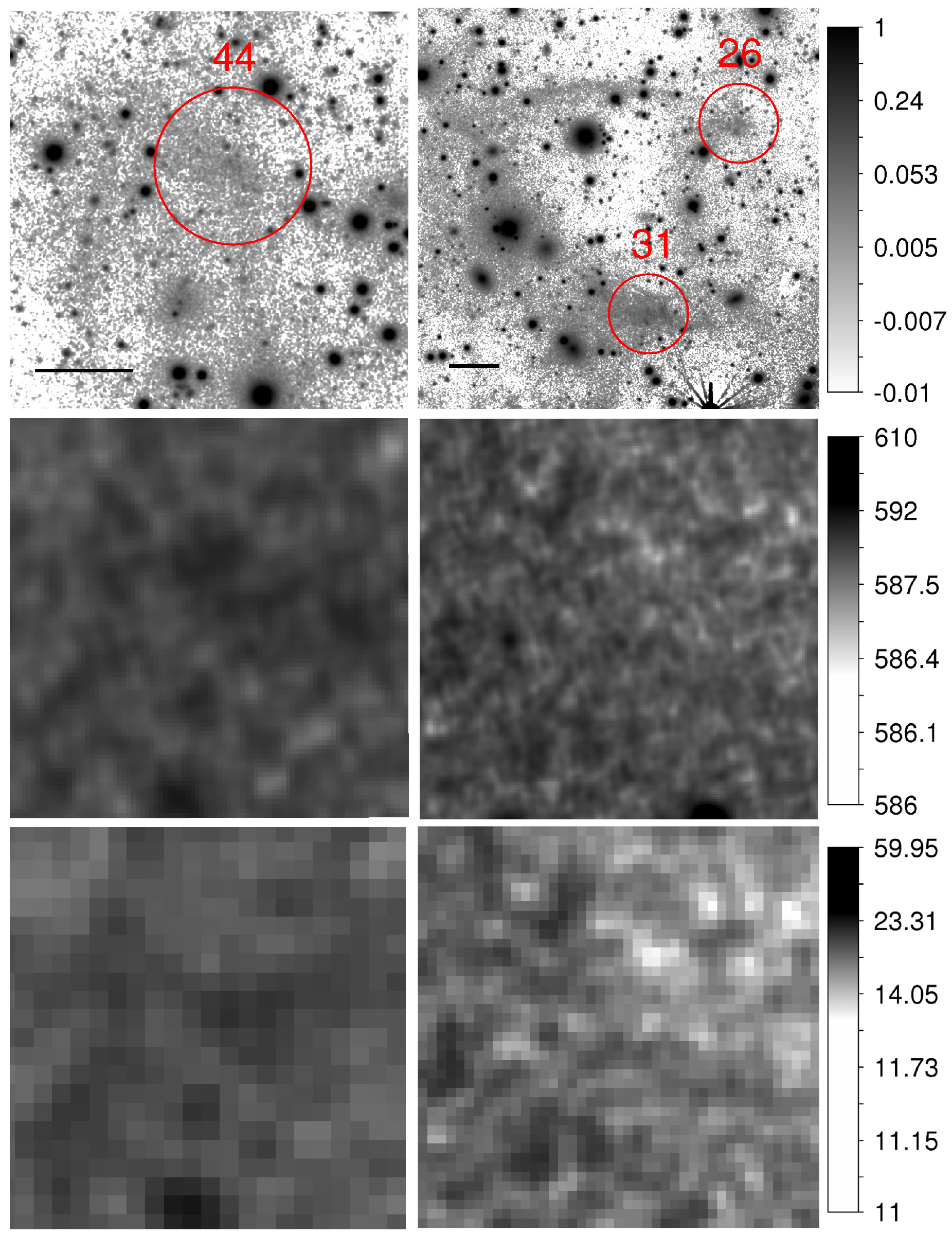}
    \caption{Objects from our sample that could be either LSB galaxies with possible tidal streams or cirrus emission. The top panels show the objects in our smoothed data, marked with red circles. The four lower panels show the corresponding regions in the {\it WISE} $12\mu\mathrm{m}$ intensity maps that trace Galactic cirrus. The original {\it WISE} intensity maps with 6\,arcsec resolution are displayed in the middle panels, the reprocessed {\it WISE} intensity maps with 15\,arcsec resolution that were cleaned from point sources are shown in the bottom panels. The height and width of the cutout regions is 2\,arcmin ($\hat{=}\,41$\,kpc) in the left-hand panels and 4\,arcmin ($\hat{=}\,81$\,kpc) in the right-hand panels, respectively. The black bar in the images in the top panels denotes a length of 10\,kpc. We see no obvious correspondence between the structures observed in our data and the $12\mu\mathrm{m}$ emission. We therefore cannot draw any firm conclusions on the nature of these structures.}
    \label{fig:fig7}
\end{figure}

In Fig.~\ref{fig:fig6}, we try to include realistic photometric uncertainties for our sample. Our major source of uncertainty in the measured total fluxes, which translate to uncertainties in half-light radii and surface brightnesses, lies in the adopted background level (see Section~\ref{sec:sec4}). To test how large the resulting uncertainties are, we probed this using inserted LSB galaxy models that were generated similarly to those described in Section~\ref{sec:sec3}. We created eight model types that span the parameter range of our sample. Four model types have $\langle\mu_V\rangle_{50} = 25.5$\,mag\,arcsec$^{-2}$, the other four have $\langle\mu_V\rangle_{50} = 26.5$\,mag\,arcsec$^{-2}$, with varying magnitudes $M_V = -12.5$ to $-15.5$\,mag and sizes $0.8 \leq r_{50} \leq 4.9$\,kpc. The models have one component S\'{e}rsic profiles with $n=1$, are nearly round (ellipticity = 0.1) and were convolved to our average seeing FWHM. We inserted 10 models of each type into one copy of our mosaic, respectively. We then measured $M_V$, $r_{50}$ and $\langle\mu_V\rangle_{50}$ similarly to our sample of real LSB candidates. We calculated the average offset between true and measured parameters for each model type, as well as the scatter of the measured parameters. 

We indicate the average parameter offsets with arrows in the right-hand panels of Fig.~\ref{fig:fig6}. The arrow tips point to the true values, with $M_V$ being systematically estimated as too faint by on average 0.4\,mag, and $r_{50}$ being underestimated by on average 0.5\,kpc. We largely preserved the true surface brightness, which results from our approach of considering the uncontaminated part of the flux profile only (see Section~\ref{sec:sec4}). The offsets in $\langle\mu_V\rangle_{50}$ are small, and do not exceed 0.1\,mag\,arcsec$^{-2}$. In general the parameter offsets are more severe for model types with the largest size and faintest surface brightness, and negligible for model types with the smallest size and brightest surface brightness. The error bars in Fig.~\ref{fig:fig6} give the standard deviation of the measured $M_V$, $r_{50}$ and $\langle\mu_V\rangle_{50}$ values for each model type, with average standard deviations of $\Delta M_V = \pm 0.3$\,mag, $\Delta r_{50} = \pm 0.3$\,kpc and  $\langle\mu_V\rangle_{50} = \pm 0.1$\,mag\,arcsec$^{-2}$.

We also tested the implications of our estimated uncertainties on our results from Section~\ref{sec:sec5.2}, and applied the average systematic offsets in $M_V$, $r_{50}$ and $\langle\mu_V\rangle_{50}$ between the models and the measured parameters of our LSB galaxy sample. In this case the number of LSB candidates in the considered parameter range of $M_V \leq -14$\,mag and $r_{50} \geq 2$\,kpc would increase to 25 candidates in the Perseus cluster core, but still only two LSB candidates would have sizes larger than $r_{50} \geq 3$\,kpc. Thus, while the number of LSB candidates would now be significantly higher in Perseus compared to the number of LSB candidates in the same parameter range in the Coma cluster core, the conclusion of only finding very few large LSB galaxy candidates in the cluster core would remain unchanged.

Since the core regions of massive clusters are characterized by a particularly high density of galaxies, one possible concern is that this may have influenced our ability of detecting large LSB galaxy candidates with $r_{50} \geq 3$\,kpc. Our tests with the inserted LSB galaxy models indicate, however, that we are in principle able to detect objects with $r_{50} > 3$\,kpc in the surface brightness range $\langle\mu_V\rangle_{50} < 27$\,mag\,arcsec$^{-2}$ in our data, if these were present (see Section~\ref{sec:sec3}). Nevertheless we might have missed objects in close vicinity to bright cluster galaxies or foreground stars, although we modelled and subtracted the light profile of the latter in most cases. The apparent absence of LSB candidates in regions around bright sources in Fig.~\ref{fig:fig4} might therefore not be a real effect.\\

Due to the location of the Perseus cluster at low Galactic latitude ($l = 13$\degr) we cannot exclude the presence of diffuse emission from Galactic cirrus in our data. Cirrus is often visible in deep wide-field imaging data, and the resulting structures can be very similar in appearance to stellar tidal streams \citep[cf.][]{Deschenes2016}. We therefore compared our candidates with possible streams to the {\it WISE}\footnote{{\it Wide-field Infrared Survey Explorer} \citep{Wright2010}} $12\mu\mathrm{m}$ data that trace Galactic cirrus, in order to search for possible counterparts in the $12\mu\mathrm{m}$ emission. Fig.~\ref{fig:fig7} shows our data in comparison to both the original {\it WISE} data with 6\,arcsec resolution, as well as to the reprocessed data from \citet{Meisner2014} with 15\,arcsec resolution that were cleaned from point sources. We clearly see diffuse emission in the $12\mu\mathrm{m}$ data at the position of Perseus. However, we are not able to identify obvious structures in the {\it WISE} maps that would match to the candidates with possible streams we observe in our data, due to the insufficient resolution of the latter. Therefore, we neither can confirm nor exclude that the nature of these structures may be cirrus emission rather than LSB galaxy candidates with tidal streams.

\section{Discussion}
\label{sec:sec6}

We detected a large number of 89 faint LSB galaxy candidates with $\langle\mu_V\rangle_{50} \geq 24.8$\,mag\,arcsec$^{-2}$ in the Perseus cluster core. It is interesting to note that all but one candidate have $r_{50} < 3$\,kpc. We thus speculate that LSB galaxies with larger sizes cannot survive the strong tidal forces in the core region and possibly have lost already a considerable amount of their dark matter content. This observation is consistent with the study of \citet{vanderBurg2016} who found a decreasing number density of faint LSB galaxy candidates in the cores of galaxy clusters. Also, the numerical simulations of \citet{Yozin2015} predicted the disruption of LSB galaxies orbiting close to the cluster centre.

The effect of tides on LSB galaxies in galaxy clusters is possibly also reflected in the radial number density distribution we observe for our sample. The nearly constant projected number density for clustercentric distances $r \geq 100$\,kpc implies that the three-dimensional distribution should actually increase with distance from the cluster centre. This may be a further argument that LSB galaxies are depleted in the cluster core region due to tidal disruption. Very close to the cluster centre, for clustercentric distances $r < 100$\,kpc, the number density drops, with only two LSB candidates from our sample being located in this region. Here tidal effects from the central cluster galaxy NGC~1275 may become apparent \citep[cf.][fig.~1]{Mathews2006}. For example, the slightly more compact peculiar galaxy SA~0426-002 ($M_B = -16.3$\,mag, $r_{50} = 2.1$\,kpc), being located only $\sim 30$\,kpc from the cluster centre, shows signs of being tidally disturbed (see Fig.~\ref{fig:fig4}, top left panel). Also, in the Fornax cluster core a drop in the number density profile of faint LSB candidates is seen within 180\,kpc of the cluster centre \citep{Venhola2017}.

We can use the observed limit in $r_{50}$ as a rough constraint on the dark matter content of the LSB candidates in the cluster centre \citep[cf.][]{Penny2009}. The tidal radius $R_{\mathrm{tidal}}$ is given by
\begin{equation}
R_{\mathrm{tidal}} = R_{\mathrm{peri}} \left( \frac{M_{\mathrm{obj}}}{M_{\mathrm{cl}}(R_{\mathrm{peri}})\,(3 + e)}\right)^{1/3},
	\label{eq:eq1}
\end{equation}
with the pericentric distance $R_{\mathrm{peri}}$, the total object mass $M_{\mathrm{obj}}$, the cluster mass $M_{\mathrm{cl}}(R_{\mathrm{peri}})$ within $R_{\mathrm{peri}}$ and the eccentricity of the orbit $e$ \citep{king1962}. We find about 50 per cent of our sample (44 objects) at projected clustercentric distances below 330\,kpc. Assuming that this is representative of the orbital pericentre for at least a fraction of the population,\footnote{While on the one hand, most objects are likely to be situated somewhat further away from the centre than the projected value suggests, on the other hand, it is also likely that their orbital pericentre is located further inwards from their current location.} we estimate $R_{\mathrm{tidal}}$ for a typical LSB candidate of our sample with $M_V = -14$\,mag and $R_{\mathrm{peri}} = 330$\,kpc, assuming an eccentric orbit with $e = 0.5$. We adopt the cluster mass profile from \citet{Mathews2006}, where $M_{\mathrm{cl}}(330\mathrm{\,kpc}) = 1.3 \times 10^{14}$\,M$_{\odot}$. 

Assuming a galaxy without dark matter, and adopting a mass-to-light ratio of $M/L_V = 2$ for an old stellar population with subsolar metallicity \citep{Bruzual2003}, the mass of an object with $M_V = -14$\,mag would be $M_{\mathrm{obj}} = 7 \times 10^7$\,M$_{\odot}$ accordingly, resulting in a tidal radius of 1.8\,kpc. This compares to a range of observed $r_{50} \simeq 1.0 - 2.5$\,kpc for LSB candidates from our sample with $M_V \simeq -14$\,mag. We note that we can generally probe our objects out to more than one half-light radius in our data, thus the tidal radius would be within the observed stellar extent. However, since most objects from our sample do not show obvious signs of current disruption, we suspect that they may contain additional mass in order to prevent tidal disruption.

If we assume a higher mass-to-light ratio of $M/L_V = 10$, the tidal radius of the same object would increase to 2.9\,kpc. For $M/L_V = 100$ the tidal radius would be $R_{\mathrm{tidal}} = 6.2$\,kpc, and for $M/L_V = 1000$ we derive $R_{\mathrm{tidal}} = 13.3$\,kpc. For $M/L_V$ close to 1000 the tidal radius is significantly larger than the observed range of half-light radii. If such a high mass-to-light ratio would be reached within the tidal radius, we might expect to find a higher number of galaxies with $r_{50} \gtrsim 3$\,kpc in the cluster core. However, {\it for $M/L_V \lesssim 100$, the tidal radius would be on the order of $1$--$2$\,r$_{50}$}, which is also consistent with the mass-to-light ratios derived from dynamical measurements of similar galaxies. For example, \citet{vanDokkum2016} found a mass-to-light ratio of $\sim 50$ within one half-light radius for one LSB galaxy in the Coma cluster ($M_V = -16.1$\,mag, $r_{50} = 4.3$\,kpc),\footnote{Based on stellar dynamics of the galaxy.} and \citet{Beasley2016a} derived a mass-to-light ratio of $\sim 100$ within one half-light radius for one LSB galaxy in Virgo ($M_g = -13.3$\,mag, $r_{50} = 2.8$\,kpc).\footnote{Based on GC system dynamics of the galaxy.} We note that based on similar analytical arguments as described above \citet{vanDokkum2015a} also estimated a dark matter fraction of $\gtrsim 100$\,per\,cent within an assumed tidal radius of $6$\,kpc for a sample of faint LSB candidates within the core region of the Coma cluster.

While the above approach gives an estimate of the radius beyond which material is likely going to be stripped, another approach to estimate the effect of tides on galaxies in clusters is to compare the density of the tidal field to the density of the orbiting galaxy \citep[cf.][]{Gnedin2003}. The density of the tidal field $\rho_{\mathrm{tidal}}$ is given by Poisson's equation, $\rho_{\mathrm{tidal}} = F_{\mathrm{tidal}} / (4 \pi G)$, where $F_{\mathrm{tidal}}$ is the trace of the tidal tensor. We consider the extended mass distribution of the cluster\footnote{Unlike in the first approach, where a point-mass approximation was used.} and approximate the strength of the tidal force at a given clustercentric distance $r_0$ as $F_{\mathrm{tidal}} = \vert dg(r)/dr\vert_{r_0}$, where $g(r)$ is the gravitational acceleration exerted by the mass of the cluster. For $g(r)$ we adopt the gravitational acceleration due to the Perseus cluster potential given by \citet{Mathews2006}, where we only consider the contribution of the NFW-profile, which is the dominant component at clustercentric distances r $\gtrsim 10$\,kpc. We approximate the average density of the orbiting galaxy, assuming spherical symmetry, as $\rho_{\mathrm{gal}} = M_{\mathrm{gal}}(R) / (4 \pi R^3 / 3)$, where $M_{\mathrm{gal}}(R)$ is the total mass of the galaxy within a radius $R$. Requiring that the density of the galaxy is larger than the tidal density to prevent its disruption, the limiting radius $R_{\mathrm{lim}}$ is given as
\begin{equation}
R_{\mathrm{lim}} \geq \sqrt[3]{\frac{3 G M_{\mathrm{gal}}(R)}{\vert dg(r)/dr\vert_{r_0}}}
	\label{eq:eq2}
\end{equation}
Considering again a typical galaxy from our sample, with $M_V = -14$\,mag at a clustercentric distance $r_0 = 330$\,kpc, we find $R_{\mathrm{lim}} = 0.8$\,kpc for $M/L_V = 2$, $R_{\mathrm{lim}} = 1.3$\,kpc for $M/L_V = 10$, $R_{\mathrm{lim}} = 2.8$\,kpc for $M/L_V = 100$ and $R_{\mathrm{lim}} = 6.1$\,kpc for $M/L_V = 1000$. Thus, in comparison to the tidal radius derived with the first approach, the limiting radius obtained with the second approach is a factor of two smaller. If we assume that $M/L_V = 100$ would be characteristic for a considerable fraction of our sample, then the limiting radius would be on the order of only $1\,r_{50}$.

Does this imply that a few of the largest LSB candidates in the Perseus cluster core should be in process of tidal disruption right now? -- We do identify three LSB candidates in Perseus that show possible signs of disruption (see panels on the right-hand side in Fig.~\ref{fig:fig4}). Candidate 44 appears to be embedded in stream like filaments. It is, however, unclear whether we see here still a bound galaxy or rather a remnant core of a stream. Candidates 26 and 31 seem to be connected via an arc-like tidal stream. This could point to a low-velocity interaction between those two candidates, since such interactions produce the most severe mass-loss. The convex shape of the stream with respect to the cluster centre might suggest that these two objects are not in orbit around the cluster centre, but instead still bound to a possibly recently accreted subgroup of galaxies. The association with a subgroup could be supported by the observation that these three candidates, together with the candidate of brighter surface brightness with tidal tails (see Fig.~\ref{fig:fig4}, lower left panel), are located closely together in a region south--west of the cluster centre, within a clustercentric distance range of $300$--$400$\,kpc. It is also interesting to note that \citet{Merritt2016} found a generally more complex and distorted morphology for LSB candidates in galaxy groups than in galaxy clusters, indicating that the group environment may play an important role in shaping galaxies of low stellar density.

The comparison to the LSB galaxy samples in Coma \citep{vanDokkum2015a,Yagi2016} showed that both cluster cores hold a similar number of faint LSB candidates with $r_{50} \geq 2$\,kpc and $M_V \leq -14$\,mag. Based on the 1.5 times lower cluster mass of Perseus\footnote{Assuming $M_{\mathrm{vir,Coma}} = 1.3\times10^{15}$\,M$_{\odot}$ \citep{Lokas2003} and $M_{\mathrm{vir,Perseus}} = 8.5\times10^{14}$\,M$_{\odot}$ \citep{Mathews2006}.}, we would expect a somewhat lower number of all galaxy types in Perseus. However, with regard to the density in the cluster core, both clusters reach a comparable galaxy surface number density within 0.5\,Mpc \citep{Weinmann2011}, thus causing comparable disruptive forces in both cluster cores. Therefore, according to the cluster mass and density, we would expect a similar or even lower number of LSB galaxies of such large size in Perseus, which is in agreement with our observations.\\

One important question to investigate would be whether there exists a possible evolutionary link between LSB galaxies that are red and quiescent and those that are blue and star-forming. The cosmological simulations of \citet{DiCintio2017} suggest that faint LSB galaxies with large sizes may form as initially gas-rich star-forming systems in low-density environments. In this context, the quenching of star formation should be related to external processes, like, e.g., ram pressure stripping. \citet{Roman2017} examined a sample of faint LSB candidates in group environments. Since they found the red LSB candidates closer to the respective group's centre than the blue systems this could imply that the group environment was efficient in removing the gas that fuels star formation. This is also seen among the dwarf galaxies of the Local Group, which show a pronounced morphology -- gas content -- distance relation \citep[see][]{Grebel2003}. However, a few quiescent and gas-poor LSB galaxies of dwarf luminosity are also observed in isolation \citep[e.g.][]{Papastergis2017}, which would not fit into this scenario. An essential aspect would be to understand whether the physical processes governing the formation and evolution of LSB galaxies are controlled by stellar density or by stellar mass. The latter could possibly explain the observed wide variety of LSB galaxy properties from low-mass dSphs to massive LSB disc galaxies.

\section{Summary and conclusions}
\label{sec:sec7}

We obtained deep $V$-band imaging data under good seeing conditions of the central regions of Perseus with PFIP at the {\it WHT} that we used to search for faint LSB galaxies in the surface brightness range of the so-called `ultra-diffuse galaxies'. We detected an abundant population of 89 faint LSB galaxy candidates for which we performed photometry and derived basic structural parameters. Our sample is characterized by mean effective surface brightnesses $24.8 \leq \langle\mu_V\rangle_{50} \leq 27.1$\,mag\,arcsec$^{-2}$, total magnitudes $-11.8 \geq M_V \geq -15.5$\,mag and half-light radii $0.7 \leq r_{50} \leq 4.1$\,kpc. A comparison to overlapping {\it HST}/ACS imaging data indicates that the sample is relatively uncontaminated by background objects.

We find good evidence for tidal disruption leading to a deficiency of LSB galaxy candidates in the central regions of the cluster. This is indicated by a constant observed number density beyond clustercentric distances of 100\,kpc and the lack of very large LSB candidates with $r_{50} \geq 3$\,kpc except for one object. However, only a few candidates show structural evidence of ongoing tidal disruption. If LSB systems are to remain gravitationally bound in the cluster core, the density limits set by the Perseus cluster tidal field require that they have high $M/L$ values of about 100, assuming a standard model for gravity.

In comparison to the Coma cluster -- with its comparable central density to Perseus -- we find that our sample statistically resembles the LSB galaxy population in the central regions of Coma. Given the same dearth of large objects with $r_{50} \geq 3$\,kpc in both cluster cores we conclude that these cannot survive the strong tides in the centres of massive clusters.

\section*{Acknowledgements}
The {\it William Herschel Telescope} is operated on the island of La Palma by the Isaac Newton Group of Telescopes in the Spanish Observatorio del Roque de los Muchachos of the Instituto de Astrof\'{i}sica de Canarias (programme 2012B/045).
We thank Simone Weinmann and Stefan Lieder for useful comments when preparing the {\it WHT} observing proposal.
CW\ is a member of the International Max Planck Research School for Astronomy and Cosmic Physics at the University of Heidelberg (IMPRS-HD).
RK gratefully acknowledges financial support from the National Science Foundation under grant no. AST-1664362.
This research has made use of the NASA/ IPAC Infrared Science Archive, which is operated by the Jet Propulsion Laboratory, California Institute of Technology, under contract with the National Aeronautics and Space Administration.




\bibliographystyle{mnras}
\bibliography{mybibliography} 








\bsp	
\label{lastpage}
\end{document}